\documentclass[12pt]{iopart}
\usepackage{mathrsfs}
\usepackage{graphicx}
\usepackage{hyperref}

\usepackage{multirow}
\usepackage{mathrsfs}

\begin{document}
\title{High-dimensional quantum state transfer in a noisy network environment}
\author{Wei Qin$^{1,2,3,4}$, Jun Lin Li$^{2,3,4}$, and Gui Lu Long$^{2,3,4}$}
\address{$^1$School of Physics, Beijing Institute of Technology, Beijing 100081, China\\
$^2$State Key Laboratory of Low-Dimensional Quantum Physics and Department of Physics, Tsinghua University, Beijing 100084,
China\\
$^3$Collaborative Innovation Center of Quantum Matter, Beijing 100084, China\\
$^4$Tsinghua National Laboratory for Information Science and Technology, Tsinghua University, Beijing 100084, China}
\ead{gllong@mail.tsinghua.edu.cn}
\begin{abstract}
We propose and analyze an efficient high-dimensional quantum state transfer protocol in an XX coupling spin network with a hypercube structure or chain structure. Under free spin wave approximation, unitary evolution results in a perfect high-dimensional quantum swap operation requiring neither external manipulation nor weak coupling. Evolution time is independent of either distance between registers or dimensions of sent states, which can enable improvements in computational efficiency. In the low temperature regime and thermodynamic limit, the decoherence by  noisy environment is studied with a model of an antiferromagnetic spin bath coupled to quantum channels via an Ising type interaction is studied. it is found that while the decoherence reduces the fidelity of state transfer, increasing intra-channel coupling can strongly suppress such effects.
These observations demonstrate that robustness of the proposed scheme.
\end{abstract}
\maketitle

\section{Introduction}
Quantum state transfer (QST) between two remote parties
is an essential ingredient of scalable quantum information processing.
Quantum channels enabling universal operations between two physically separated
registers have been proposed in a variety of quantum systems, including superconducting transmission lines
\cite{superconductor},
Coulomb coupling trapped ions \cite{ion}, and optical photons \cite{photon1,photon2}.
In particular, coupled quantum spin systems have
attracted much attention for short distance quantum communication in recent decades
\cite{spin1,spin2,spin3,spin4,spin5,spin6,spin7}.
Such coupled-spin quantum channels provide an interesting
alternative to  direct register interactions and interfacing with photonic flying qubits.
For this reason, most of the feasible proposals have been proposed for perfect QST,
in which the quantum channels commonly rely upon engineered couplings entailing
suitable dispersion relations \cite{prespin1,prespin2,prespin3},
weak couplings undergoing an effective Rabi oscillation or a ballistic
regime \cite{weakspin1,weakspin2,weakspin3}, and dynamical manipulation
\cite{ctrspin1,ctrspin2,ctrspin3,ctrspin4}.

By contrast to two-dimensional quantum systems,
i.e., qubits, high-dimensional quantum systems serving as qudits have
been extensively studied due to their key advantages in large capacity and
high security.
While such advantages have been explored in quantum communication
ranging from quantum key distribution to
quantum teleportation
\cite{qudit_commu1,qudit_commu2,qudit_commu3,qudit_commu4,qudit_commu5,qudit_commu6},
qudits systems are also candidates for quantum computation ranging from
universal quantum simulations and asymptotically optimal quantum circuits to
one-way quantum computation and
fault-tolerant quantum computation
\cite{qudit_compu1,qudit_compu2,qudit_compu3,qudit_compu4,qudit_compu5,qudit_compu6,qudit_compu7,qudit_compu8,qudit_compu9}.
Indeed, several high-dimensional QST protocols have been presented
in coupled-spin chains \cite{qudit_spin1,qudit_spin2,qudit_spin3}.

Similar to classical supercomputers involving parallel processing \cite{hypercube_supercumputer},
hypercube topology can potentially allow for diverse applications in
QST \cite{prespin1,hypercube_transfer1,hypercube_transfer2,hypercube_transfer3},
quantum random walks \cite{hypercube_walk1,hypercube_walk2} and
quantum search algorithms \cite{hypercube_search1,hypercube_search2}. Hypercube
is a generalization from three-dimensional cube to high-dimensional configuration, and
it possesses many appealing topological features, e.g., node-symmetry, good balance
between nodes and edges \cite{hypercube_supercumputer}.
Prior state transfer schemes in hypercubes have focused on qubits using
either spins \cite{prespin1}, single-photons \cite{hypercube_transfer2},
or multiphoton states \cite{hypercube_transfer3}.
In this article, we propose an efficient high-dimensional QST
in XX coupling spin hypercube networks. The hypercubes are multiple Cartesian products
of a chain of either two or three spins, which works as building blocks
of these topologies \cite{Graph_Theory}. Upon performing free spin wave approximation,
unitary evolution for a specific time
results in a perfect high-dimensional swap gate and
evolution time is
independent of either distance between two registers or dimension of sent state \cite{hypercube_transfer3}.
It does not require weak coupling, projective measurement, external modulation, and even coupling engineering. In addition,
utilizing Schwinger picture yields a perfect mirror high-dimension swap operation in a coupled-spin chain of
arbitrary length.
Numerical results confirm that the spin-wave interaction leads to the
leakage of quantum information, and under the free spin wave approximation,
the perfect high-dimensional QST is achieved.

The proposed efficient high-dimensional QST occurs in closed quantum channel.
However, a quantum channel can rarely be isolated from its surrounding environment, especially
the spin bath. Thus it is necessary to discuss the decoherence effects on such protocols.
The considered decoherence is characterized by a pure dephasing model for quantum channels
coupled to an antiferromagnetic (AFM) spin bath via a typical Ising interaction
\cite{spinbath1,spinbath2}, allowing for the conservation of channel energy. In low temperature regime,
unitary evolution analytically gives a swap gate experiencing decoherence. To be specific, we perform numerical simulations when
the spin bath is in the thermal equilibrium state. We find that while increasing either the spin bath temperature or
the bath-channel coupling enhances the irreversible leakage of quantum information
into the spin bath,
strong intrachannel coupling can depress the decoherence effects to
ensure the high fidelity of state transfer. Observing these can help us to understand the
decoherence effects on quantum communication in coupled quantum spin systems.

The paper is organized as follows. In Sec. \ref{se:section2}, we calculate the Hamiltonian and
the operator evolution under free spin wave approximation.
In Sec. \ref{se:section3}, we study the high-dimensional QST
in hypercubes and chains, respectively.
In Sec. \ref{se:section4}, specifically, the decoherence of
spin bath in the thermal equilibrium state
is studied. The last section is a summary.

\section{Model and calculation}
\label{se:section2}

\begin{figure}[!ht]
\begin{center}
\includegraphics[width=8.5cm,angle=0]{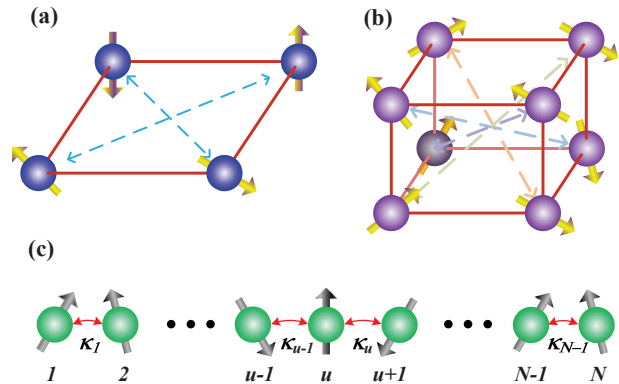}
\caption{Schematic configurations for two-fold (a) and three-fold (b)
Cartesian products of a two-spin chain with uniform coupling strength $\kappa$.
Unitary evolution for a time $\tau_{1}=\pi/\left(4S_{0}\kappa\right)$ results in a
high-dimensional swap operation between two spins along each main diagonal.
In a coupled-spin chain of length $N$ (c), unitary evolution for a time
$\tau'=\pi/\left(2S_{0}g_{0}\right)$ enables
a mirror-like high-dimensional swap operation with respect to
its center through engineering inter-spin coupling strengths
$\kappa_{u}=g_{0}
\sqrt{u\left(N-u\right)}$.}\label{fig1}
\end{center}
\end{figure}
The composite system being considered consists of a quantum spin-$S_{0}$ network
which enacts as a quantum channel,
and an AFM spin-$S$ bath whose lattice is divided into two
identical sublattices $a$ and $b$. The network Hamiltonian with spins
coupled to their nearest neighbors on a finite lattice of site $N_{0}$ is of XX coupling
\begin{equation}\label{eq:network_XX_coupling}
H_{S}=\sum_{u,v=1}^{N_{0}}K_{uv}S_{0,u}^{-}S_{0,v}^{+},
\end{equation}
where $K$ is a $N_{0} \times N_{0}$ coupling matrix
and $K_{uv}$ represents the coupling strength
between sites $u$ and $v$, $S^{\pm}=S^{x}\pm iS^{y}$
are two ladder operators and $S^{\mu}$ $\left(\mu=x,y,z\right)$
is the $\mu$ component of a spin operator $\mathbf{S}$.
The AFM spin bath Hamiltonian is described by a Heisenberg model
\begin{equation}
H_{B}=\frac{J}{2}\sum_{\mathbf{i},\mathbf{\delta}}\mathbf{S}_{a,\mathbf{i}}\cdot
\mathbf{S}_{b,\mathbf{i}+\mathbf{\delta}}
+\frac{J}{2}\sum_{\mathbf{j},\mathbf{\delta}}\mathbf{S}_{b,\mathbf{j}}\cdot
\mathbf{S}_{a,\mathbf{j}+\mathbf{\delta}},
\end{equation}
where $J>0$ is the exchange interaction,
$\mathbf{\delta}$ is a vector joining a site to its nearest neighbor,
$\mathbf{S}_{a}$ and $\mathbf{S}_{b}$
are spin operators in the sublattices $a$ and $b$, respectively.
The interaction Hamiltonian between the network and the spin bath is characterized
by a typical Ising model \cite{spinbath2}
\begin{equation}
H_{I}=-\frac{J_{0}}{\sqrt{N}}\sum_{u=1}^{N_{0}}S_{0,u}^{z}\otimes\sum_{\mathbf{i}}
\left(S_{a,\mathbf{i}}^{z}+S_{b,\mathbf{i}}^{z}\right),
\end{equation}
with the coupling strength $J_{0}$ and the number of spins $N$ in each magnetic sublattice. The dynamics of
composite system is governed by the total Hamiltonian $H=H_{S}+H_{B}+H_{I}$, and
upon introducing the Holstein-Primakoff (HP) transformation \cite{HPT},
\begin{eqnarray}\label{eq4}
&&S^{-}_{a,i}=a^{\dag}_{i}\sqrt{2S-a^{\dag}_{i}a_{i}},\quad S^{z}_{a,i}=S-a^{\dag}_{i}a_{i},\nonumber\\
&&S^{-}_{b,j}=\sqrt{2S-b^{\dag}_{j}b_{j}}b_{j},\quad S^{z}_{b,j}=b^{\dag}_{j}b_{j}-S,\nonumber\\
&&S^{-}_{0,u}=c^{\dag}_{u}\sqrt{2S_{0}-c^{\dag}_{u}c_{u}},\quad S^{z}_{0,u}=S_{0}-c^{\dag}_{u}c_{u},
\end{eqnarray}
the total Hamiltonian can be expressed in terms of boson operators,
wherein conservation of total spin $z$ projection of channel, $\left[\sum_{u=1}^{N_{0}}
S_{0,u}^{z},H\right]=0$, becomes conservation of boson number, $\left[\sum_{u=1}^{N_{0}}
c_{u}^{\dag}c_{u},H\right]=0$.

The network is initialized to a simple ferromagnetic order with spins aligning in a
parallel way. The low-lying level states of a sender, ranging from $|0\rangle$ to $|d-1\rangle$,
are employed to encode quantum
information as an input state, and we assume that the dimension $d$ of sent state is much
smaller than $2S_{0}$, $d\ll2S_{0}$. The conservation law ensures that $\langle c_{u}^{\dag}c_{u}
\rangle\ll2S_{0}$, and the HP transformation is simplified to $S_{0,u}^{-}=\left(2S_{0}\right)^{1/2}c_{u}^{\dag}$,
which leads to
\begin{equation}\label{eq:reduced_Channel_H}
H_{S}=2S_{0}\sum_{u,v=1}^{N_{0}}K_{uv}c_{u}^{\dag}c_{v}.
\end{equation}
The subsequent diagonalization of this tight-binding Hamiltonian is realized through
an orthogonal transformation $Q$, such that $\Lambda=QKQ^{\dag}$ with $\Lambda_{qq'}=\lambda_{q}
\delta_{qq'}$. This transformation results in $H_{S}=\sum_{q=1}^{N_{0}}\varepsilon_{q}
c_{q}^{\dag}c_{q}$, where $c_{q}=\sum_{u=1}^{N_{0}}Q_{qu}c_{u}$ and $\varepsilon
_{q}=2S_{0}\lambda_{q}$.

The spin bath surrounding the quantum channel
is in the low-temperature and low-excitation limit such that the spin operators
can be approximated as $S_{a,\mathbf{i}}^{-}=\left(2S\right)^{1/2}a_{\mathbf{i}}^{\dag}$
and $S_{b,\mathbf{j}}^{-}=\left(2S\right)^{1/2}b_{\mathbf{j}}$ \cite{SWA}. Working with Fourier transformation and
Bogoliubov transformation yields
\begin{equation}
H_{B}=E_{0}+\sum_{\mathbf{k}}\omega_{\mathbf{k}}\left\{
\left(\alpha_{\mathbf{k}}^{\dag}\alpha_{\mathbf{k}}+\frac{1}{2}\right)
+\left(\beta_{\mathbf{k}}^{\dag}\beta_{\mathbf{k}}+\frac{1}{2}\right)\right\},
\end{equation}
where $\alpha_{\mathbf{k}}$ and $\beta_{\mathbf{k}}$ are two degenerate
AFM mangnon branches, and $E_{0}=-z_{0}JNS(S+1)$. The spin-wave elementary excitation spectra (EES) is
\begin{equation}
\omega_{\mathbf{k}}=z_{0}JS\sqrt{1-\gamma_{\mathbf{k}}^{2}},
\end{equation}
wherein $z_{0}$ is the number of the nearest neighbors, and
$\gamma_{\mathbf{k}}=z_{0}^{-1}\sum_{\mathbf{\delta}}e^{i\mathbf{k}
\cdot \mathbf{\delta}}$ is the structure factor.
In the low-energy regime, corresponding to long wavelengths, $k\ll l$, the EES is
linear: $\omega_{k}=\left(2z_{0}\right)^{1/2}JSkl$, with the lattice constant $l$.
The interaction Hamiltonian is likewise transformed to
\begin{equation}
H_{I}=-\frac{J_{0}}{\sqrt{N}}\sum_{q=1}^{N_{0}}\left(S_{0}-c_{q}^{\dag}c_{q}\right)
\otimes \sum_{\mathbf{k}}\left(\beta_{\mathbf{k}}^{\dag}\beta_{\mathbf{k}}
-\alpha_{\mathbf{k}}^{\dag}\alpha_{\mathbf{k}}\right).
\end{equation}

The Heisenberg equation drives an operator evolution in Heisenberg picture,
$c_{m}^{\dag}\left(t\right)=e^{iHt}c_{m}^{\dag}\left(0\right)e^{-iHt}$. By applying
this equation and the commutation relation $\left[H_{I},H\right]=0$,
\begin{equation}
c_{m}^{\dag}\left(t\right)=e^{iY\left(t\right)}\tilde{c}_{m}^{\dag}\left(t\right).
\end{equation}
Here, the quantum noise operator $Y\left(t\right)$ reads
\begin{equation}
Y\left(t\right)=\frac{J_{0}t}{\sqrt{N}}\sum_{\mathbf{k}}\left(\beta_{\mathbf{k}}^{\dag}\beta_{\mathbf{k}}
-\alpha_{\mathbf{k}}^{\dag}\alpha_{\mathbf{k}}\right),
\end{equation}
which arises from the AFM spin bath, and
$\tilde{c}_{m}^{\dag}\left(t\right)=
e^{iH_{S}t}c_{m}^{\dag}\left(0\right)e^{-iH_{S}t}$
is the free evolution of network. Owing to $c^{\dag}_{q}\left(t\right)=
e^{i\varepsilon_{q}t}c^{\dag}_{q}$, and
\begin{equation}
\left(e^{i2S_{0}Kt}\right)_{um}=\sum_{q=1}^{N_{0}}Q^{\dag}_{uq}e^{i\varepsilon_{q}t}
Q_{qm};
\end{equation}
thus, $\tilde{c}_{m}^{\dag}\left(t\right)$ turns out to be
\begin{equation}\label{eq:free_evolution}
\tilde{c}_{m}^{\dag}\left(t\right)=\sum_{u=1}^{N_{0}}
\left(e^{i2S_{0}Kt}\right)_{um}c_{u}^{\dag},
\end{equation}
and the free evolution of network is determined by its coupling matrix.

\section{Perfect high-dimensional quantum state transfer}
\label{se:section3}

In this section, high-dimensional QST protocols with high fidelity
in hypercubes and chains are proposed, respectively. We begin with
the case of hypercubes, which are multiple Cartesian products of either two-spin chain
$G_{1}$ or three-spin chain $G_{2}$ working as basic building blocks to build such
coupled-spin hypercubes. The coupling matrix
of a hypercube network $G$ in Eq. (\ref{eq:network_XX_coupling}) is $K=\kappa A\left(G\right)$,
and $\kappa$ is the coupling strength between two spins in the network, which
is characterized by its adjacency matrix $A\left(G\right)$ \cite{Graph_Theory}.

After $g$-fold Cartesian products of either of the two simple chains,
$A\left(G\right)$ is given by
\begin{equation}
A\left(G\right)=\sum_{j=0}^{g-1}I^{\otimes j}\otimes A\left(G_{\theta}\right)\otimes
I^{\otimes \left(g-j-1\right)},
\end{equation}
where $I$ is an identity matrix, and $A\left(G_{\theta}\right)$ are adjacency matrices
of $G_{\theta}$ for $\theta=1,2$. As a consequence,
\begin{equation}
e^{i2S_{0}Kt}=\left[e^{i2S_{0}\kappa A\left(G_{\theta}\right)t}\right]^{\otimes g},
\end{equation}
and it results in
\begin{equation}\label{eq:kronecker_product}
\left(e^{i2S_{0}K\tau_{\theta}}\right)_{um}=i^{\theta g}\delta_{u,N+1-m}
\end{equation}
at the following evolution time
\begin{equation}
\tau_{\theta}=\pi/\left(2^{1+1/\theta}S_{0}\kappa\right).
\end{equation}
Substituting Eq. (\ref{eq:kronecker_product}) into Eq. (\ref{eq:free_evolution}) leads to
\begin{equation}
\tilde{c}_{m}^{\dag}\left(\tau_{\theta}\right)=i^{\theta g}c_{N+1-m}^{\dag}.
\end{equation}
Upon absorbing a phase factor of $i^{\theta g}$ into $c_{N+1-m}^{\dag}$,
unitary evolution under $H_{S}$ for a time $\tau_{\theta}$ results in
$\tilde{c}_{m}^{\dag}\left(\tau_{\theta}\right)=c_{N+1-m}^{\dag}$, which
allows for the realization of a swap operation between two spins $m$ and
$N+1-m$. It is determined by the natural dynamics of Hamiltonian and requires
neither external modulation nor inter-spin coupling strength engineering.
Moreover, the evolution time
is independent of distance between the two remote spins, and the speedup of
perfect high-dimensional QST is possible.
For simplicity, we take the square and cube networks as examples, as shown
in Fig. \ref{fig1}(a) and (b),  which are two-fold
and three-fold Cartesian products of $G_{1}$, respectively,
the time evolution builds a swap gate between two spins
along each main diagonal at the optimal time.

While the case of a multi-dimensional hypercube has been chosen to focus on, we extend
such results to a one-dimensional coupled-spin system. By choosing
\begin{equation}
K_{uv}=\kappa_{u-1}\delta_{u,v+1}+\kappa_{u}\delta_{u,v-1},
\end{equation}
in the Hamiltonian of Eq. (\ref{eq:network_XX_coupling}), such that it provides
a Hamiltonian of a coupled-spin chain with coupling strengths $\kappa_{u}$ between
two spins $u$ and $u+1$, as shown in Fig.
\ref{fig1}(c). The coupling matrix is identical to a pseudo Hamiltonian $H'=g_{0}L_{x}$,
where $L_{x}$ is the $x$ component
of a fictitious spin $L=\left(N_{0}-1\right)/2$, and $g_{0}$ is a coupling parameter.
In Schwinger picture, $L_{x}$ can
be rewritten in terms of two boson operators as \cite{Schwinger_picture}
\begin{equation}
L_{x}=\frac{1}{2}\left(l^{\dag}_{1}l_{2}+l_{1}l_{2}^{\dag}\right),
\end{equation}
and hence, $H'$ is the Hamiltonian governing two bosons with coupling strength $g_{0}/2$.
Unitary evolution under $2S_{0}H'$ for a time $\tau'=\pi/\left(2S_{0}g_{0}\right)$ gives
$l_{1}^{\dag}\left(\tau'\right)=il_{2}^{\dag}$ and $l_{2}^{\dag}\left(\tau'\right)=il_{1}^{\dag}$,
yielding
\begin{equation}
\left(e^{i2S_{0}K\tau'}\right)_{um}=i^{N_{0}-1}\delta_{u,N+1-m}.
\end{equation}
Therefore, $\tilde{c}_{m}\left(\tau'\right)$ is evaluated as
\begin{equation}
\tilde{c}_{m}\left(\tau'\right)=i^{N_{0}-1}c_{N_{0}+1-m}^{\dag}.
\end{equation}
Upon absorbing a factor of $i^{N_{0}-1}$ into $c_{N_{0}+1-m}^{\dag}$, as desired, it
enables a swap operation between spins $m$ and $N_{0}+1-m$,
$\tilde{c}_{m}\left(\tau'\right)=c_{N_{0}+1-m}^{\dag}$, and a mirror inversion
of quantum states with respect to the center of the chain is implemented.

\begin{figure}[!ht]
\begin{center}
\includegraphics[width=7.5cm,angle=0]{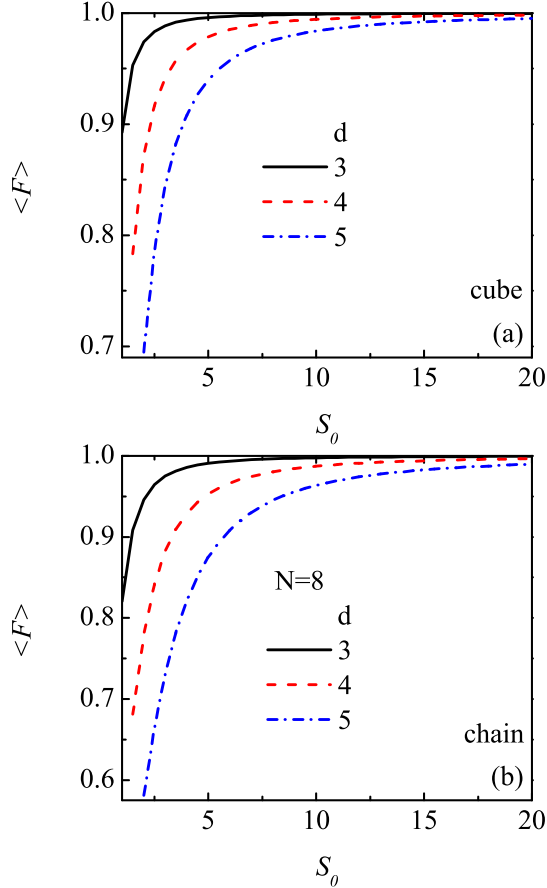}
\caption{The average fidelity $\langle F\rangle$ as a function of quantum spin number
 with either $d=3$ for solid black lines, $d=4$ for dashed red lines, or $d=5$ for
 dotted-dashed blue lines. (a) Coupled-spin cube; (b) Coupled-spin chain of length $N=8$.
 Here, the evolution time is the optimal time for the two different configurations.}\label{fig2}
\end{center}
\end{figure}

To confirm perfect high-dimensional QST, we perform numerics,
as shown in Fig. \ref{fig2}. The initial state of network is supposed
to be
\begin{equation}\label{eq:network_initial_state}
|\Psi\rangle_{ini}=|\varphi\rangle_{m}\otimes |0\rangle_{\bar{m}},
\end{equation}
where $|\varphi\rangle_{m}=\sum_{\nu=0}^{d-1}C_{\nu}|\nu\rangle$ is the sent state at the spin $m$,
and $|0\rangle_{\bar{m}}$ represents that all spins are in the vacuum state apart from just
one spin at the site $m$. The normalized coefficients $C_{\nu}$ can be measured by the Hurwitz
parametrization with $d-1$ polar angles $\chi_{p}$ and $d-1$ azimuthal angles $\vartheta_{p}$ as \cite{HDM,Lihui}
\begin{eqnarray}\label{hur}
\Bigg(\cos\vartheta_{d-1},\sin\vartheta_{d-1}\cos\vartheta_{d-2}e^{i\chi_{d-1}},&
\sin\vartheta_{d-1}\sin\vartheta_{d-2}
\cos\vartheta_{d-3}e^{i\chi_{d-2}},\nonumber\\
&\cdots,\prod_{i=1}^{d-1}\sin\vartheta_{i}e^{i\chi_{1}}\Bigg),
\end{eqnarray}
where $0\leq \vartheta_{p} \leq \pi/2$ and $0\leq \chi_{p} < 2\pi$ for $p=1,2,...,d-1$.
The average fidelity over the whole manifold of such states is
\begin{equation}
\langle F\left(t\right) \rangle=\frac{1}{V_{d}}\int_{CP^{d-1}} F\left(t\right) dV.
\end{equation}
Here, $F\left(t\right)$ is defined by $F\left(t\right)=_{1}\langle\varphi|\rho_{m}\left(t\right)|\varphi\rangle_{1}$
with $\rho_{m}\left(t\right)$ is the reduced density matrix of spin $m$
at the evolution time $t$.
The volume element on
the generalized Bloch sphere of Eq. (\ref{hur}) in the complex space $CP^{d-1}$ is
\begin{equation}
dV=\prod^{d-1}_{p=1}\cos\vartheta_{p}(\sin\vartheta_{p})^{2p-1}d\vartheta_{p}d\chi_{p},
\end{equation}
and the total volume of the $2(d-1)$-dimensional manifold of pure states is $V_{d}=\pi^{d-1}/(d-1)$.

The average fidelity varies as a function of spin number $S_{0}$, as shown in
Fig. \ref{fig2}. The infidelity, $\epsilon=1-\langle F\rangle$, results from the leakage of
quantum information into the
spin wave interaction found in nonlinear terms of HP transformation.
In such regime $2S_{0}\sim d$, evolution of quantum channel is dominated
by the spin wave interaction suppressing the average fidelity.
By ensuring, $2S_{0}/d\ll1$, the spin wave interaction
is negligible and evolution is governed by the free spin wave found in linear terms of
HP transformation, enabling a
perfect high-dimensional QST.

\section{Spin bath in the thermal equilibrium state}
\label{se:section4}

To be specific, we consider a spin bath in the thermal equilibrium state,
wherein its variables
are distributed in an uncorrelated thermal equilibrium mixture of states,
and density matrix satisfies the Boltzmann distribution
\begin{equation}
\rho_{B}=\frac{1}{Z}e^{-H_{B}/T},
\end{equation}
where $Z=\mathrm{Tr}\left(e^{-H_{B}/T}\right)$ is a partial function and $T$ represents the
temperature. In combination with Eq. (\ref{eq:network_initial_state}), this yields
an initial state of composite system, including quantum channel and spin bath,
as a direct product
\begin{equation}
\rho\left(0\right)=\left(|\varphi\rangle\langle\varphi|\right)_{m}
\otimes \left(|0\rangle\langle0|\right)_{\bar{m}}\otimes \rho_{B}\left(0\right).
\end{equation}
Evolution for a time $\tau'$ drives the state to $\rho\left(\tau'\right)=e^{-iH\tau'}\rho\left(0\right)e^{iH\tau'}$,
and the reduced density matrix of quantum channel is $\rho_{S}\left(\tau'\right)=\mathrm{Tr}_{\mathrm{B}}\left[
\rho\left(\tau'\right)\right]$. Without loss of generality, we have chosen a couple-spin chain,
it is directly analogous in the case of hypercube quantum channels.
Upon utilizing $|\nu\rangle=\left(c^{\dag}\right)^{\nu}/\sqrt{\nu!}|0\rangle$ and
$e^{-iH\tau'}c^{\dag}\left(0\right)e^{iH\tau'}=c^{\dag}\left(-\tau'\right)$, the finial state of spin
$N_{0}+1-m$ is calculated as
\begin{equation}
\rho_{N_{0}+1-m}\left(\tau'\right)=\sum_{\nu,\nu'=0}^{d-1}\rho_{\nu\nu'}\left(\tau'\right)
D_{\nu\nu'}\left(T\right),
\end{equation}
with $\rho_{\nu\nu'}\left(\tau'\right)=C_{\nu}C_{\nu'}^{*}|\nu\rangle\langle \nu'|$, and
a decoherence factor $D_{\nu\nu'}\left(T\right)$ given by
\begin{equation}
D_{\nu\nu'}\left(T\right)=\frac{1}{Z}\mathrm{Tr}_{B}\left[e^{-i\left(\nu-\nu'\right)
Y\left(\tau'\right)-H_{B}/T}\right],
\end{equation}
which turns out to be $D_{\nu\nu'}\left(T\right)=r_{\nu\nu'}^{+}r_{\nu\nu'}^{-}/r_{0}^{2}$. Therein,
\begin{equation}
r_{0}=\prod_{\mathbf{k}}\frac{1}{1-e^{-\omega_{\mathbf{k}}/T}},
\end{equation}
and
\begin{figure}[!ht]
\begin{center}
\includegraphics[width=8.0cm,angle=0]{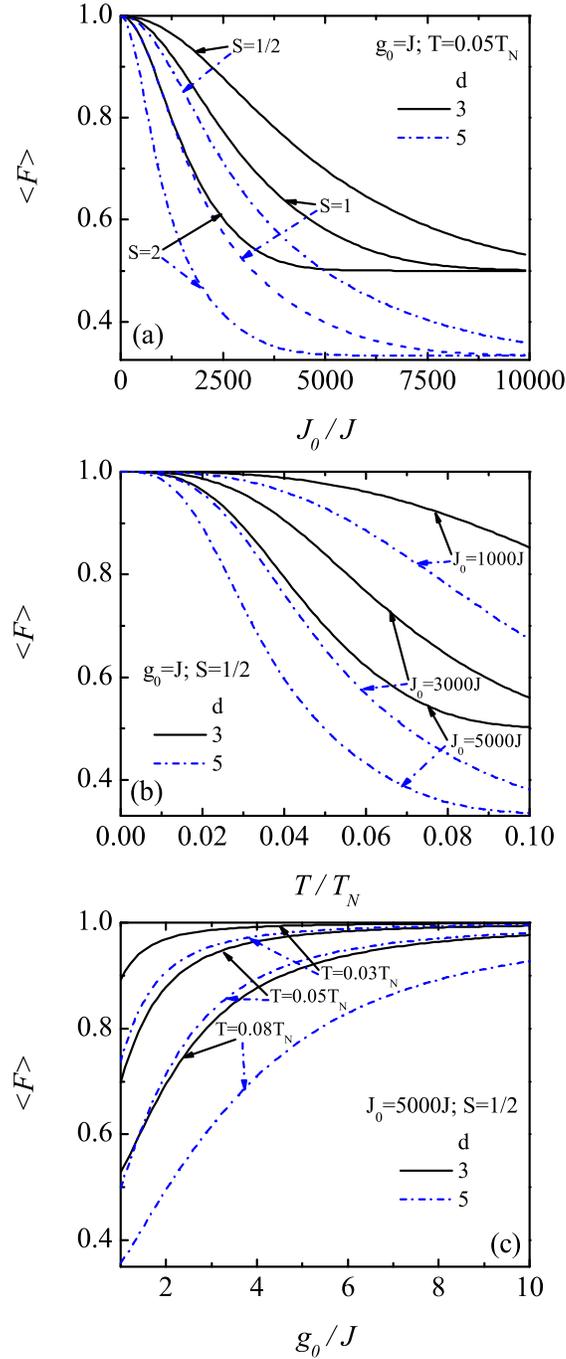}
\caption{The decoherence effects on the high-dimensional QST proposal in a coupled-spin chain
with either $d=3$ for solid black lines, or $d=5$ for dotted-dashed blue lines.
The average fidelity as functions of (a) $J_{0}$ with $g_{0}=J$ and $T=0.05T_{N}$,
(b) $T$ with $g_{0}=J$ and $S=1/2$, and (c) $g_{0}$ with $J_{0}=5000J$ and $S=1/2$.}\label{fig3}
\end{center}
\end{figure}
\begin{equation}
r_{\nu\nu'}^{\pm}=\prod_{\mathbf{k}}\frac{1}{1-e^{\pm \phi_{\nu\nu'}}e^{-\omega_{\mathbf{k}}/T}},
\end{equation}
where $\phi_{\nu\nu'}=\left(\nu-\nu'\right)J_{0}\tau' N^{-1/2}$. Upon introducing quantities
\begin{equation}
\xi_{\pm}=\frac{f_{\pm}\left(\phi_{\nu\nu'}\right)}{\phi_{\nu\nu'}^{2}},
\end{equation}
and
\begin{equation}
f_{\pm}\left(\phi_{\nu\nu'}\right)=\int_{0}^{\infty}x^{2}\ln\frac{1-e^{\pm i\phi_{\nu\nu'}}
e^{-\omega_{\mathbf{k}}/T}}{1-e^{-\omega_{\mathbf{k}}/T}}dx,
\end{equation}
with $x=kl$. The decoherence factor is transformed to
\begin{equation}
D_{\nu\nu'}\left(T\right)=e^{-\left(\xi_{+}+\xi_{-}\right)\left(\nu-\nu'\right)^{2}J_{0}^{2}\tau'^{2}
/\left(2\pi^{2}\right)},
\end{equation}
where the sum over $\mathbf{k}$ has been replaced by an integral
\begin{equation}
\sum_{\mathbf{k}}=\frac{V}{\left(2\pi\right)^{3}}\int dk 4\pi k^{2},
\end{equation}
with $N=V/l^{3}$. In the thermodynamic limit, $N\rightarrow \infty$,
it leads to $\xi_{0}=\xi_{\pm}$, and
\begin{equation}
\xi_{0}=\frac{\pi^{2}T^{3}}{12z_{0}\sqrt{2z_{0}}J^{3}S^{3}}.
\end{equation}
Thus the decoherence factor becomes
\begin{equation}
D_{\nu\nu'}\left(T\right)=e^{-\left(\nu-\nu'\right)^{2}\tau'^{2}/\tau_{c}^{2}},
\end{equation}
with a decoherence time
\begin{equation}
\tau_{c}=\frac{\pi}{J_{0}\sqrt{\xi_{0}}}.
\end{equation}
In combination, the fidelity of spins $m$ and $N_{0}+1-m$ is
\begin{equation}
F\left(\tau'\right)=\sum_{\nu,\nu'=0}^{d-1}|C_{\nu}|^{2}|C_{\nu'}|^{2}D_{\nu\nu'}\left(T\right).
\end{equation}

To illustrate the decoherence effects of spin bath on the quantum channels, we perform numerics,
as shown in Fig. \ref{fig3}.
Under the random phase approximation, the N\'{e}el temperature of an
AFM system reads \cite{SWA}
\begin{equation}
T_{N}=\frac{z_{0}JS\left(S+1\right)}{3\zeta},
\end{equation}
with $\zeta=N^{-1}\sum_{\mathbf{k}}\left(1-\gamma_{\mathbf{k}}\right)^{-1}$.
The summation over $\mathbf{k}$ is restricted in the first Brillouin Zone of sublattice,
such that it has half the volume of atomic Brillouin Zone, yielding $\zeta=1.51638$ for simple
cubic lattice and $\zeta=1.39320$ for body-center cubic lattice.
Owing to the validity of spin wave theory only in low temperature regime,
the spin bath temperature is restricted below $0.1T_{N}$, $T\leq0.1T_{N}$.
The average fidelity varies as a function of either coupling strength, $J_{0}$, or
spin bath temperature, $T$, as shown in Figs. \ref{fig3}(a) and \ref{fig3}(b).
The infidelity, $\epsilon$, results from the leakage of quantum information into the spin bath.
Increasing either $J_{0}$ or $T$ enhances such irreversible process.
Fig. \ref{fig3}(c) plots the average fidelity as a function of coupling parameter,
$g_{0}$. Strong coupling parameter can partly counteract the
thermal effects to prevent the leakage of quantum information into the spin bath and
ensure the high fidelity.
~\\

\section{Summary}
\label{se:section5}

In this paper, we have studied an high-dimensional QST protocol in
spin networks of either hypercubes or chains coupled to an antiferromagentic spin bath.
Under the free spin wave approximation and in ideal conditions without decoherence,
time evolution presents a perfect high-dimensional QST between two
registers of arbitrary distance for both of the two configurations, requiring neither weak coupling,
external modulation nor projective measurement.
The essence of our method is that a perfect swap operation is allowable for a chain of either two or three bosons.
One of its direct applications is it can potentially provide the high-dimensional entanglement
distribution for quantum computation. Moreover, state transfer independent of either the
distance between two registers or the dimension of sent state enables the speedup of
computational efficiency.
In the low temperature regime and thermodynamic limit,
the decoherence induced by the AFM spin bath has been investigated in this work. Increasing
either the spin bath temperature or the channel-bath coupling, the decoherence becomes more serious.
However, strong intrachannel coupling can partly counteract the decoherence effects, and ensure
the high fidelity to demonstrate robustness against external noise.
Observing these will deepen
us understanding of the decoherence effects on quantum communication in the coupled quantum
spin systems.

While we have focused on the specific case of a hypercube topology, the conceptual framework
can be used in a wide range of topologies by means of the
free spin wave approximation to yield the tight-binding Hamiltonian described in
Eq. (\ref{eq:reduced_Channel_H}). By ensuring $\left(e^{i2S_{0}K\tau}\right)_{um}=\delta_{u,N_{0}+1-m}$
for a specific evolution time $\tau$, the achievement of a perfect high-dimensional QST is possible.

\section{Acknowledgements}
\label{se:section6}
This work was supported by the National Natural Science Foundation
of China under Grant Nos. 11175094 and 91221205, the National Basic
Research Program of China under Grant No.
2011CB9216002. GLL also thanks the support of Center of
Atomic and Molecular Nanoscience of Tsinghua University.

\end{document}